\documentclass[amssymb,aps,prl,reprint,nofootinbib,showpacs,superscriptaddress,groupedaddress]{revtex4-2}  
\usepackage{latexsym}
\usepackage{amsmath}
\usepackage[utf8x]{inputenc}
\usepackage[T1]{fontenc}
\usepackage{subdepth}
\usepackage{color}
\usepackage{soul}
\usepackage{csquotes}
\usepackage{amsthm}
\usepackage{amssymb}
\usepackage{tensor}
\usepackage{dsfont}
\usepackage{tikz-cd}

\begin{document}
\title{Chern-Simons Inspired  Massive Gravity}
\author{J. Chagoya$^1$}
\author{M. Sabido$^{2}$}
\author{A. Silva-Garc\'ia$^{2}$}
\affiliation{
$^{1}$Unidad Acad\'emica de F\'isica, Universidad Aut\'onoma de Zacatecas,
98060, M\'exico.\\
$^{2}$Departamento  de F\'{\i}sica de la Universidad de Guanajuato, A.P. E-143, C.P. 37150, Le\'on, Guanajuato, M\'exico.
 }%
\begin{abstract}
In this work we propose a new  2+1 theory of gravity. We start with a modification of Chern-Simons 2+1 gravity. By introducing a vector field $\Phi$ to the usual composition of the vierbein and connection, we derive a new action that includes the Einstein-Hilbert term, a cosmological constant and a polynomial term for the vector field. This allows us to write a Chern-Simons inspired  Massive Gravity. After solving the polynomial function, the action can be rewritten as a Born-Infeld type action. Moreover, we show that the second order expansion of this action is equivalent to New Massive Gravity.
We explore  solutions to the theory and show that it  has a BTZ solution with an effective cosmological constant in terms of the parameters of the theory,  global Lifshitz  space-times as well as regular Lifshitz black holes, and asymptotically flat black holes.

 \end{abstract}

\maketitle

{\it Introduction. ---} One of the most successful playgrounds to understand the theoretical intricacies of general relativity is 2+1 gravity. For example, the BTZ solution \cite{PhysRevLett.69.1849}  has been fundamental in shedding light on  the properties of black holes. Moreover, 2+1 gravity has many remarkable features such as renormalizability, integrability, topological invariance, etc., that guided the quest for a 4-dimensional quantum theory of gravity. The origin of these is connected to the fact that 2+1 gravity can be written as a Chern-Simons gauge theory \cite{Witten:1988hc}, allowing to treat 2+1 gravity in a similar way to Yang-Mills theory. 
Following this line of reasoning, in \cite{Garcia-Compean:1999fdl}
an (anti)self-dual formulation of 2+1 gravity is proposed, which allows to write the {\it ``standard''} and {\it ``exotic''} actions as linear combinations of the self-dual and anti-self-dual actions. Furthermore, 2+1 gravity has been derived in the context of Topological M-theory \cite{Dijkgraaf:2004te,Chagoya:2016zhy} and used in understanding the entropy of exotic BTZ black holes \cite{Chagoya:2020uog,Townsend:2013ela}. Moreover, an interesting feature of the Chern-Simons action is that its quantization is well known and consequently  2+1 gravity is a quantizable theory. 

Interest in 2+1 gravity was reignited by new massive gravity (NMG) \cite{Bergshoeff:2009hq}. This theory is a generalization of 2+1 general relativity that includes a propagating massive spin-2 particle with helicity $\pm 2$. Also, it was argued that topological massive gravity (TMG) can be considered a special case of NMG, as in the linearized limit has the equivalent field equations to NMG. This can be seen from the factorization of the spin 2 Fierz-Pauli equation in 3 dimensions, because it is equivalent to linearized TMG \cite{Bergshoeff:2009hq,Dalmazi:2008zh,Deser:2002iw}. 
Black hole solutions were explored in NMG \cite{Bergshoeff:2009aq}, in particular the presence of BTZ and Lifshitz solutions was derived~\cite{PhysRevD.90.044026}. Extensions to NMG have been proposed, for example a Born-Infeld generalization of NMG \cite{Gullu:2010pc}, or more recently a non local version has been derived from Einstein Gravity \cite{Bueno:2023dpl}.
Considering that NMG is a generalization of 2+1 GR and that 2+1 GR can be written as a Chern-Simons theory, one can entertain the idea that NMG can have a similar construction. For example, it is well known that 3+1 GR can be written as a Yang-Mills type theory and a particular construction is the MacDowell-Mansouri formulation of gravity \cite{MacDowell:1977jt}. This theory and in particular the self-dual generalization \cite{PhysRevD.50.R3583}, has been used to understand theoretical aspects, like the Immirzi parameter \cite{Mercuri:2010yj, Obregon:2012zz} and noncommutative gravity \cite{Garcia-Compean:2003nix, Garcia-Compean:2002dgp}. But more recently, a generalization of this formulation, was used to obtain a
Yang-Mills-type derivation for a vector-tensor gravity contained in generalized Proca theories~\cite{Chagoya:2022wba}.
From this discussion, it seems natural to look 
for a generalization of 2+1 gravity starting from its Chern-Simons construction~\cite{Garcia-Compean:1999fdl} and then considering the approach in \cite{Chagoya:2022wba} in order to obtain a new theory after a particular identification of the tetrad. As we will see, this new theory has an action analogous to  the triple master action in \cite{Bergshoeff:2009hq}. This new action has a cubic polynomial constraint from which a new 2+1 massive gravity is derived. In particular, if we consider up to the quadratic terms on the constraint, the resulting action is equivalent to NMG. 
This leads to the main result of this paper: a Chern-Simons inspired new massive gravity. This is achieved by extending the action that results from the Chern-Simons formulation, including a free coupling parameter for every order in the constraint. Finally we study stationary solutions,  obtaining a BTZ solution with an effective cosmological constant that can be set to zero for appropriate values of the parameters. We also show that CSIMG admits asymptotically flat BH solutions for the full theory, this is in contrast to NMG, where flat solutions are only present in the quadratic theory. We also obtain a new exact regular Lifshitz-like black hole solutions. 

{\it Chern-Simons Inspired Massive Gravity. --- } 
NMG  is a unitary theory that contains a  massive spin-2 particle, it can be derived from the Lagrangian
\begin{eqnarray}\label{NMGf}
    S[g_{\mu\nu},f_{\mu\nu}]&=&\int d^3x \sqrt{-g}\left[ R+2\bar\lambda+f^{\mu\nu}G_{\mu\nu}\right.\\
    &+& \left .\frac{1}{4}m^2\left(f^{\mu\nu}f_{\mu\nu}-f^2\right)\right],\nonumber
\end{eqnarray}
where $f_{\mu\nu}$ is an auxiliary symmetric tensor field with trace  $f=g^{\mu\nu}f_{\mu\nu}$. It was showed that after expanding around a Minkowski space time, one gets the Fierz-Pauli theory for massive spin-2 particles. Solving for the auxiliary field $f_{\mu\nu}$ and substituting in the action, we get 
\begin{equation}\label{NMGaction}
    S=\int d^3x \sqrt{-g}\left[ R+2\bar\lambda+\frac{1}{m^2}\left(R_{\mu\nu}R^{\mu\nu}-\frac{3}{8}R^2\right)\right],
\end{equation}
which is the originally proposed version of NMG \cite{Bergshoeff:2009hq}.\\
Since 2+1 gravity can be written  as a gauge theory with a Chern-Simons action \cite{Witten:1988hc}, as an alternative to NMG, we can try to introduce  degrees of freedom in the  2+1  Chern-Simons construction and  relate these additional degrees of freedom with the massive gravity theories (\textit{e.g.},  \cite{Bergshoeff:2009hq}).
Let us start with the Chern-Simons action \cite{Garcia-Compean:1999fdl} 
 \begin{align}\label{sdIpm}
    &{I}_{CS} =\\
    &\int \epsilon^{\mu\nu\rho}\left({ A_\mu}^{
    AB}{} \partial_\nu{A_{\rho AB}}{}+\frac{2}{3}{ A_\mu{}^B}_A{}{A_\nu{}^C}_B{}{A_\rho{}^A}_{C}{} \right),\nonumber
\end{align}
where $A,B,C=0,1,2,3$ and $\eta_{AB}=diag(-1,1,1,1)$. After the identifications
 $({A_\mu}^{3a}, {A_\mu}^{ab})=(\sqrt{\lambda}{e_\mu}^{a},{\omega_\mu}^{ab})$, one gets  the standard action for 2+1 Chern-Simons gravity.
 It was showed in \cite{Witten:1988hc} that this action is equivalent to the 2+1 Einstein-Hilbert action with cosmological constant. For $\lambda>0$ (de Sitter) one has the symmetry group SO(3,1), and for $\lambda<0$ (anti-de Sitter) one has $SO(2,2)$.
Following the approach in \cite{Chagoya:2022wba}, we can introduce the new degrees of freedom to the 2+1  Chern-Simons gravity action.
We start by imposing a shift only on the  $A^{3a}$ component, 
$A^{3a}\rightarrow e^a+\Phi^a$,  while maintaining the same connection, consequently the curvature is the same as in Chern-Simons 2+1  gravity.\footnote{We assume that the triad is
torsionless, and consequently  $\Phi^a$ cannot be removed by a redefinition.} Therefore, the connections are upgraded to  {${A_\mu}^{AB}=(\sqrt{\lambda}(e_\mu{}^a+\Phi_\mu{}^a), \epsilon^{abc}\omega_{\mu c})$}.  
When we apply this connection into  the action Eq.\eqref{sdIpm} we obtain a brand new modified action, 
\begin{align}\label{CSmod}
    &I_\Phi=I_{CS}\\
    &+\int \epsilon^{\mu\nu\alpha}\left[\Phi_{\mu a} \left(\partial_\nu{\omega_\alpha}^{a}-\partial_\alpha{\omega_\nu}^{a}+\epsilon^a_{~bc}{\omega_\nu}^b{\omega_\alpha}^c\right)\right.\nonumber\\
    &+\left.\frac{\lambda}{3}\epsilon_{abc}\left(3e_\mu{}^a e_\nu{}^b\Phi_\alpha{}^c+3e_\mu{}^a\Phi_\nu{}^b\Phi_\alpha{}^c+\Phi_\mu{}^a\Phi_\nu{}^b\Phi_\alpha{}^c\right)\right].\nonumber
\end{align}
To write  the action  in terms of the three dimensional space-time Riemann curvature tensor, we consider a 3-dimensional space-time $M$ with metric $g_{\mu\nu}$ and an orthonormal frame field $e$. 
In this frame the standard action takes the form
\begin{align}\label{grI}
I_\Phi&= \int d^3 x \sqrt{-g}\left[2\lambda+ R -2G^{\alpha  \mu}\Phi _{\alpha  \mu  }\right.\\
&+\left.\lambda\left( 2\Phi-\epsilon^{\mu\nu\alpha}\epsilon_{\gamma\delta\alpha}\Phi_\mu^\gamma\Phi_\nu^\delta-\tfrac{1}{3}  \epsilon^{\mu\nu\alpha}\epsilon_{\gamma\delta\beta} \Phi_\mu^\gamma\Phi_\nu^\delta\Phi_\alpha^\beta \right)\right],\nonumber
\end{align}
where $\Phi_{\mu\nu}$ is an auxiliary symmetric tensor field with trace  $\Phi=g^{\mu\nu}\Phi_{\mu\nu}$. Because this action is the analog of Eq.(\ref{NMGf}), one can consider the  Chern-Simons result as the starting point to propose a more general action. In some sense this is similar to the case of four dimensional massive gravity, where for particular values of the coupling constants, one can have theories with different behaviors \cite{Lu:2011zk}. Therefore, lets us propose a Chern-Simons inspired massive gravity, introducing independent couplings to each of the terms that depend only on $\Phi$,
\begin{eqnarray}\label{nueva}
S_\Phi&=& \int d^3 x \sqrt{-g}\left[2 \lambda  + R -2G^{\alpha  \mu}\Phi _{\alpha  \mu  }+2\zeta \Phi \right .\\
&-&\left .\sigma \epsilon^{\mu\nu\alpha}\epsilon_{\gamma\delta\alpha}\Phi_\mu^\gamma\Phi_\nu^\delta- \tfrac{1}{3}  \xi\epsilon^{\mu\nu\alpha}\epsilon_{\gamma\delta\beta} \Phi_\mu^\gamma\Phi_\nu^\delta\Phi_\alpha^\beta \right].\nonumber
\end{eqnarray}
which we will refer to as CSIMG. We can establish the connection with NMG by
considering up to quadratic terms in $\Phi$,
\begin{eqnarray}
  I^{(2)}_\Phi&=&  \int d^3 x \sqrt{-g} \left[2 \lambda  + R   -2 G^{\alpha  \mu  } \Phi _{\alpha  \mu  }\right .\\
   &+&\left.  \sigma \left(2\frac{\zeta}{\sigma}\Phi + \Phi ^2 - \Phi _{\alpha  \mu  } \Phi ^{\alpha  \mu  }\right)\right],\nonumber
\end{eqnarray}
solving for $\Phi$ we can rewrite the action in terms of the curvature. After an appropriate reparametrization of the constants, the action takes the form of Eq.(\ref{NMGaction}),
the action of NMG, with the effective cosmological constant $\Lambda$  written in terms of $\lambda$ and $\xi$.

Now we consider  the full CSIMG action with up to cubic terms. Its variation with respect to $\Phi$ gives
\begin{align}\label{eq:vphi1}
    G_{\mu  \nu  }  =& \zeta  g_{\mu  \nu  } +\sigma  \left(g_{\mu  \nu  } \Phi ^{\alpha  }{}_{\alpha  } -  \Phi _{\mu  \nu  }\right) \\
    &- \xi  \left(\tfrac{1}{2} g_{\mu  \nu  } \Phi _{\alpha  \beta  } \Phi ^{\alpha  \beta  }  - \tfrac{1}{2} g_{\mu  \nu  } \Phi ^{\alpha  }{}_{\alpha  } \Phi ^{\beta  }{}_{\beta  }\right .\nonumber\\
     &+ \left .\Phi ^{\alpha  }{}_{\alpha  } \Phi _{\mu  \nu  }  - \Phi _{\mu  }{}^{\alpha  } \Phi _{\nu  \alpha  }\right). \nonumber
\end{align}
In order to get rid of the linear terms of $\Phi$, we make $\Phi_{\mu\nu}=-\sigma g_{\mu\nu}/\xi+\Omega_{\mu\nu}/\xi$, and after the substitution in~\eqref{eq:vphi1} one gets
\begin{equation}
    G_{\mu  \nu  }  +\left(\frac{\sigma^2}{\xi}-\zeta\right) g_{\mu  \nu  } =-\frac{1}{ 2\xi} \epsilon_{\mu\alpha\beta}\epsilon_{\nu\delta\gamma}\Omega^{\alpha\delta}\Omega^{\beta\gamma}.
\end{equation}
Note that  the right-hand side in the last line is the adjoint of the tensor $\Omega$. Solving the previous equation one gets
\begin{align}\label{eq:omemunu}
  \Omega_{\mu\nu}&=
   \pm \sqrt{\det(\xi  {G^\mu}_{\nu} + (\sigma ^2  - \zeta  \xi ) {g^\mu}_{\nu})}\nonumber\\
    &\times\left[\xi  G_{\mu\nu} + (\sigma ^2 - \zeta  \xi ) g_{\mu\nu}\right]^{-1}.
\end{align}
The solution for $\Phi$ is found by plugging this into $\Phi_{\mu\nu}=-\sigma g_{\mu\nu}/\xi+\Omega_{\mu\nu}/\xi$.
After substituting in the original action we obtain a Born-Infeld type action \cite{Beltr_n_Jim_nez_2018},
     \begin{align}\label{LrelBI}
      & I= \int d^3 x \sqrt{-g}\left[ \left(1  - \frac{\sigma }{\xi }\right) R+2 \lambda + \frac{4 \sigma ^3}{\xi ^2}  - \frac{6 \zeta  \sigma }{\xi }\right. \nonumber\\
       &\left.\pm4\xi^{-2}\sqrt{(\sigma^2-\zeta\xi)^3}\sqrt{\det\left(  {g^\mu}_{\nu}+\frac{\xi  {G^\mu}_{\nu}}{\sigma^2-\zeta\xi}\right)}\right].
   \end{align}
If we consider the expansion of the determinant to second order and
after redefining the constants we recover the NMG action in Eq.(\ref{NMGf}).

{The correspondence to the Chern-Simons formulation  is found by setting $\sigma=\zeta=\xi=\lambda$,  following the same steps as before the resulting action is
\begin{align}
  I=&\frac{4}{\sqrt{\lambda}}\int dx^3 \sqrt{-g}\left[\sqrt{\det(G^\mu{}_\nu)}\right].
\end{align}
}
{\it Static Solutions ---} To find solutions we can follow two approaches: solve the equations of motion in the determinant formulation Eq.\eqref{LrelBI}, or we can solve for the action in Eq.\eqref{nueva}, where we solve for  $\Phi$ and the metric. As Lifshitz solutions are not too common and are one of the interesting features of NMG, we will focus our interest not only on BTZ black holes but more importantly on Lifshitz black hole solutions.

For 2+1 gravity the first solution was found in \cite{PhysRevLett.69.1849}, this kind of solution is known as BTZ black hole. To explore BTZ-like solutions, we propose the line element,
 \begin{equation}\label{BTZmetric}
     ds^2=-N(r)^2dt^2+\frac{dr^2}{N(r)^2}+r^2\left(d\phi+N_\phi dt\right)^2.
 \end{equation}
Substituting the metric into the EOM for the action Eq.(\ref{LrelBI}), we find two solutions for the function $N(r)$,
\begin{align}\label{J0BTZ3}
  N_{\pm}{}^2=-M+\Lambda_{\pm} r^2+\frac{J^2}{4r^2},
\end{align}
with
\begin{align}\label{eq:btzcond}
&\Lambda_{\pm}=\frac{2 \zeta ^2 \xi -\zeta  \sigma  (3 \xi +\sigma )+\lambda  \xi ^2-\lambda  \xi  \sigma +2 \sigma ^3}{(\sigma -\xi )^2}\\
&\pm\frac{2 \sqrt{\left(\sigma ^2-\zeta  \xi \right)^2 \left(\zeta ^2-\zeta  (\xi +\sigma )+\lambda  \xi -\lambda  \sigma +\sigma ^2\right)}}{(\sigma -\xi )^2}\nonumber.
\end{align}
For the BTZ solution, $\Lambda_{\pm}$ is an effective cosmological constant given in terms of the parameters $\zeta$, $\sigma$ and $\xi$.\\
{The Chern-Simons case has the asymptotically flat solution 
\begin{equation}\label{flat}
     ds^2=-\left(\mu+ b r\right)dt^2+\frac{dr^2}{\mu+b r}+r^2d\phi^2.
 \end{equation}
For the Chern-Simons  case  the Ricci scalar is not present in the action, this is the same situation as in NMG, where Eq.(\ref{flat}) is a solution to only the quadratic term. In the full CSIMG,
for the non-rotating black hole $(J=0)$ the special case $
0=\zeta ^2-\zeta  (\xi +\sigma )+\lambda \left( \xi - \sigma\right) +\sigma ^2$
implies  $\Lambda_{+}=\Lambda_{-}=\Lambda$, and allows the existence of a solution given by 
\begin{equation}\label{metric}
     ds^2=-\left(\mu+ b r+\Lambda r^2\right)dt^2+\frac{dr^2}{\mu+b r+ \Lambda  r^2}+r^2d\phi^2,
 \end{equation}
 where
\begin{align}
    \Lambda=&\frac{(2 \zeta -\lambda -\sigma ) \left(\zeta ^2-\lambda  \sigma \right)}{(\zeta -\sigma )^2}.
\end{align}
This solution is not valid for arbitrary values of the free parameters. After substituting Eq.(\ref{metric}) in the field equations, we get  the condition
\begin{align}
     B(r)\pm\sqrt{B(r)^2}=0
\end{align}
where
\begin{align}
 &B(r)=\frac{\left(\zeta ^2-\lambda  \sigma \right)}{(\zeta -\lambda ) (\zeta -\sigma )^3} 
\bigg[(\zeta -\sigma )^2 \left(\zeta ^2-\zeta  \sigma +\sigma  (\sigma -\lambda )\right) \nonumber\\
 &\left.b +2 r (\zeta -\lambda ) \left(\zeta ^2-\lambda  \sigma \right)^2\right].
\end{align}
Hence, in general, the solution exist for $B(r)<0$ for the positive branch or $B(r)>0$  for the negative one.  In particular, if $b>0$ the solution exist
if  $0<\zeta <\sigma$, and $0<\lambda \leq\frac{\zeta ^2}{\sigma } $ for the positive\footnote{In the case of the negative branch for the solution of $\Phi_{\mu\nu}$, the parameters satisfy $0<\zeta <\sigma$ and $\zeta ^2/\sigma \leq\lambda<\zeta$.} branch of $\Phi_{\mu\nu}$.   

The Ricci curvature scalar is 
\begin{align}
R=-6\Lambda-\frac{2 b}{r},
\end{align}
meaning that the  scalar curvature is asymptotically constant and has  a singularity at  the origin (the same applies for $R_{\mu\nu}R^{\mu\nu}$ and $R_{\mu\nu\alpha\beta}R^{\mu\nu\alpha\beta}$ the other curvature scalars). We get an  asymptotically flat solution provided that $2 \zeta= \lambda +\sigma$. This condition does not remove the Ricci term from the action and therefore, unlike NMG, there are asymptotically flat solutions for the complete theory.}

For New massive Gravity, given by Eq.\eqref{NMGaction}, Lifshitz black hole solutions have been studied  in \cite{Oliva_2009}.  These solutions emerge by proposing the line element 
\begin{equation}\label{lifshitzmetric}
ds^2=-\left(\frac{r}{l}\right)^{2z}\Big(1-H(r)\Big)dt^2+\frac{l^2}{r^2}\frac{dr^2}{\Big(1-H(r)\Big)}+\frac{r^2}{l^2}d\phi^2,
\end{equation}
where $z$ is a dynamic exponent and $l$ is a scale parameter.
 A broad analysis of the spectra of $z$ values is shown in \cite{PhysRevD.90.044026} and the only valid solutions meeting the boundary conditions occur for $z=1$ and $z=3$.

For the action Eq.\eqref{nueva} the situation is different. Studying first the case of a global Lifshitz metric ($H(r)=0$), one notices that the equations of motion are considerably simpler if one fixes the constants $\zeta=\xi$, $\sigma=-\xi$. Then, one finds that the Lifshitz solution exists for any $z$ under the conditions\footnote{Taking the negative sign in Eq.\eqref{eq:omemunu}, otherwise one would need $l<0$.}
\begin{align}\label{flat_lif}
\lambda + \xi -\frac{z^{3/2}}{l^3 \sqrt{\xi}}  & = 0,\\  
(1-z+z^2)\xi - 2 l z^{1/2} \xi^{3/2} & =0.\nonumber
\end{align}
Solutions without fixing $\zeta$  and $\sigma$ do exist as well; however, the resulting expressions for $\lambda$ and $\xi$ are too cumbersome. These solutions will be presented and analyzed in detail elsewhere. Here, we keep  $\zeta=\xi$, $\sigma=-\xi$ and proceed to  study Lifshitz-like black hole solutions in the special case $z=3$, which is of interest for the purpose of comparing with the NMG $z=3$ Lifshitz black hole. Unlike NMG, for the action Eq.\eqref{nueva} an asymptotic ($r\to\infty$) analysis shows that we need to relax the metric ansatz Eq.\eqref{lifshitzmetric} to the more general form
\begin{equation}\label{liflikemetric}
ds^2=-\left(\frac{r}{l}\right)^{2z}\Big(1-F(r)\Big)dt^2+\frac{l^2}{r^2}\frac{dr^2}{\Big(1-F(r)\Big)}+\frac{r^2}{l^2}d\phi^2,
\end{equation}
where both $H$ and $F$ are required to vanish as $r\to\infty$. Then, an exact solution is found as
\begin{align}
1-F(r)& = \left(1-\frac{l^2 \mu }{r^2}\right) \left(1+\frac{2 \mu  l^2}{r^2}\right)^2, \\
1 - H(r) & = 1- \frac{l^2 \mu }{r^2}.\nonumber
\end{align}
The function $H(r)$ -- and therefore the black hole horizon -- coincides with the solution for the $z=3$ Lifshitz black hole in NMG. Assuming $\mu>0$, the additional factor in $F(r)$ does not lead to new roots of $g_{tt}$, however; it has important consequences for the Ricci curvature, which turns out to be everywhere regular (assuming again $\mu>0$),
\begin{equation}
R= -\frac{26}{l^2}+\frac{48 \mu}{2 \mu  l^2+ r^2}.
\end{equation}
As $r\to 0$ we have $R\to -2/l^2$, while for $r\to\infty$ we get $R\to-26/l^2$. Similarly, the curvature scalars $R_{\mu\nu}R^{\mu\nu}$ and $R_{\mu\nu\alpha\beta}R^{\mu\nu\alpha\beta}$ are everywhere regular.

Another case of interest for comparison with NMG is $z=1$. In this case the solution for the metric becomes the static ($J=0$) BTZ black hole but with different constraints on the parameters of the model than those arising for the same solution NMG. This traces back to the assumptions $\zeta=\xi$, $\sigma=-\xi$, which makes the theory necessarily different from NMG, where we would need $\xi=0$ but $\sigma\neq 0$. In our case, the constraint is
$\xi=-\lambda+2/l^2$, while in NMG one needs $l^2 \lambda = 1 - 1/(4 l^2 \sigma) $. Notice that $\xi=0$ implies $\lambda = 2/l^2$, which is the condition for obtaining the BTZ solution in $2+1$ Einstein gravity. Finally and as expected, this solution is the special case $J=0, \zeta=\xi, \sigma=-\xi$ of Eq.(\ref{J0BTZ3}) and Eq.(\ref{eq:btzcond}).

{\it Discussion ---}  Starting from the Chern-Simons action, with an appropriate identification of the gauge fields (a vector is introduced together with the tetrad) and the introduction of free parameters, a generalization of NMG is proposed. A similar idea was presented in \cite{delPino:2015mna} where they introduce and anti-symmetric contorsion to the spin connection. In our case, the formulation is based on the algebraic structure of the Lagrangian of the original theory in the version involving an auxiliary field. This construction ensures that the scalar and vector sectors of the theory are non-dynamical, indeed the scalar sector of the new terms in the Lagrangian reduces to a total derivative. The equations of motion for the full Lagrangian involve at most second order derivatives of both the metric and the auxiliary field, and the dynamics of the auxiliary field come purely from the coupling to the Einstein tensor already present in NMG. No new degrees of freedom are added with respect to NMG, the resulting theory can be considered as an extension of Fierz-Pauli for a massive spin-2 field in $2+1$-dimensions. 
By integrating out the auxiliary field, we obtain a Born-Infeld type action for the cubic theory and if we expand the determinant to second order we get NMG.

{One of the interesting features of NMG is the existence of the asymptotically Ads black hole 
as well as Lifshitz black hole solution. Asymptotically flat solutions in 2+1 gravity are not common, these are not present in 2+1 GR, and for NMG are only present for the pure quadratic theory \cite{Oliva:2009ip}. Moreover, in Non-linear Massive Gravity (NLMG) \cite{Bueno:2023dpl} asymptotically flat solutions are only conjectured.} 
The theory  presented in this work, as in NMG, admits a BTZ solution with a cosmological constant. This cosmological constant is written in terms of the parameters of the theory and can be made to vanish, thus unlike NMG, in CSIMG we found asymptotically flat solutions for the full theory and not only for the pure quadratic theory. Also, this theory
admits Lifshitz solutions.  The simplest solution is for global Lifshitz space-times, which exists for any $z$ that satisfies Eq.(\ref{flat_lif}). Regarding Lifshitz black holes, we find a novel solution in the particular case $z=3$ and with a specific choice of the coupling constants, that represents an asymptotically AdS black hole. The horizon of this BH is the same as for the Lifshitz BH solution in NMG with $z=3$, but, unlike the NMG case, for this solution $g_{rr}\neq g^{-1}_{tt}$; the extra term in $g_{tt}$ introduces modifications such that the curvature scalars are regular at $r=0$. Therefore, CSIMG has a regular Lifshitz black hole solution for $z=3$. Solutions with other choices of the coupling constants will be presented in detail elsewhere. 
The question of solutions with other values of $z$ remains open.

The startling differences between the solutions of CSIMG and NMG motivates the search of other solutions that have such an interesting behavior. We can ask ourselves if there are gravitational solitons or wormholes. These ideas are being explored and will be reported elsewhere. {It would also be interesting to explore the implications for the AdS/CFT duality. For instance, NMG has been re-derived from a holographic $c$-theorem and a Lagrangian involving up to four derivative curvature invariants, and it has been shown that a sixth order Lagrangian is also compatible with such theorem~\cite{Sinha:2010ai}. Given that CSIMG admits a closed form in terms of curvatures, Eq.\eqref{LrelBI}, one could analyze the existence of a $c$-theorem for this theory; this would also call for the inclusion of a matter term.} 


{\it Acknowledgments ---}This work is  supported by  CONAHCYT grant  DCF-320821. 
\bibliographystyle{unsrt}
\bibliography{ref}
\end{document}